\documentclass[conference]{IEEEtran}
\IEEEoverridecommandlockouts
\usepackage{cite}
\usepackage{amsmath,amssymb,amsfonts}
\usepackage{algorithmic}
\usepackage{graphicx}
\usepackage{textcomp}
\usepackage{xcolor}
\usepackage{float}

\newcommand{\unit}[1]{\mbox{$\rm \,#1$}}
\newcommand{\Pst}{P_{st}}
\newcommand{\uct}{u_{c}\left(t\right)}
\newcommand{\umodt}{u_\mathrm{mod}\left(t\right)}

\newcommand{\dUU}{\left( \Delta U / U \right) }
\newcommand{\fc}{f_{c}}
\newcommand{\fm}{f_{m}}
\newcommand{\const}{\mathrm{const}}
\newcommand{\boldfc}{\pmb f_{c}}
\newcommand{\mc}{m_{c}}

\def\BibTeX{{\rm B\kern-.05em{\sc i\kern-.025em b}\kern-.08em
    T\kern-.1667em\lower.7ex\hbox{E}\kern-.125emX}}
\begin{document}
	
\IEEEoverridecommandlockouts
\IEEEpubid{\begin{minipage}{\textwidth}\ \\[10pt]
		\centering\footnotesize{\newline \newline \newline \newline \copyright 2022 IEEE. Personal use of this material is permitted. Permission from IEEE must be obtained for all other uses, in any current or future media, including reprinting/republishing this material for advertising or promotional purposes, creating new collective works, for resale or redistribution to servers or lists, or reuse of any copyrighted component of this work in other works. DOI:~10.1109/ICHQP53011.2022.9808552}
\end{minipage}}

\title{IEC Flickermeter Measurement Results for Distorted Modulating Signal while Supplied with Distorted Voltage
\thanks{This research was funded in whole or in part by National Science Centre, Poland -- 2021/41/N/ST7/00397, the Foundation for Polish Science (FNP) -- Stipend START 45.2021, the Ministry of Education and Science -- 0212/SBAD/0573. For the purpose of Open Access, the author has applied a~CC--BY public copyright licence to any Author Accepted Manuscript ({AAM}) version arising from this submission.}
}

\author{\IEEEauthorblockN{Piotr~Kuwa{\l{}}ek}
\IEEEauthorblockA{\textit{Institute of Electrical Engineering and Electronics} \\
\textit{Poznan University of Technology}\\
Poznan, Poland \\
piotr.kuwalek@put.poznan.pl}

}

\maketitle

\begin{abstract}
The paper presents IEC flickermeter measurement results for voltage fluctuations modelled by amplitude modulation of distorted supply voltage. The supply voltage distortion caused by electronic and power electronic devices in the ``clipped cosine" form is assumed. This type of supply voltage distortion is a common disturbance in low voltage networks. Several arbitrary distorted waveforms of the modulating signal with different modulation depth and modulating frequency up to approx.~1\unit{\textbf{kHz}} are selected to determine the dependence of severity of voltage fluctuation on their shape. The paper mainly presents the dependence of voltage fluctuation severity with a frequency greater than 3$\boldfc$, where $\boldfc$ is the power frequency. The voltage fluctuation severity and the dependencies associated with it have been determined on the basis of numerical simulation studies and experimental laboratory tests.
\end{abstract}


\section{Introduction}
The most common power quality disturbances are voltage fluctuations and voltage distortions~\cite{b1}, especially considering low voltage (LV) networks. In LV networks, ``clipped cosine" voltage distortion~\cite{b2,b3,b4} most often occurs, which is the result of power electronic devices effects, in particular input stages of switching power supplies. Considering the common use of switching power supplies, the supply voltage in LV networks is almost always distorted~\cite{b1}. In addition, voltage fluctuations occur due to load changes in real power grids. Taking into account indicated cases, voltage fluctuations and voltage distortions often occur simultaneously. The effect of simultaneous occurrence of indicated power quality disturbances, can be an increase in the voltage fluctuation severity caused by disturbing loads changing their state with a frequency greater than~3$\fc$~\cite{b5}, where $\fc$~is the power frequency. It is worth noting that so far only voltage fluctuations caused by disturbing sources with a frequency of up to~$3\fc$ have been considered~\cite{b6,b7,b8,b9}, because a sinusoidal supply voltage has been considered. However, such (pure sine) supply voltage rarely occurs in real power grids, especially in LV networks. The proven possibility of increase in the voltage fluctuation severity with a frequency greater than~3$\fc$~\cite{b5} makes it necessary: 
\begin{itemize}
	\item modification of currently used measuring instruments~\cite{b10,b11,b12}, or use of other measures of voltage fluctuation severity assessment~\cite{b13,b14,b15,b16};
	\item comprehensive analysis and diagnostics of voltage fluctuations, considering potential voltage fluctuations sources (including power electronic devices switching with high-frequency, or groups of loads causing voltage fluctuations with a resultant frequency greater than~3$\fc$);
	\item modification of methods of identification (recognition) and localization of disturbing sources, including sources of voltage fluctuations with a frequency greater than~3$\fc$ (currently, the most common methods in the literature are methods of identification and localization disturbing loads changing their operating state with a frequency to~$\fc$~\cite{b17,b18,b19,b20,b21}; however, there are also solutions considering loads changing their state with a frequency of up to~3$\fc$~\cite{b22,b23,b24}).
\end{itemize}

Taking the above into account, the paper presents the analysis of voltage fluctuations with a frequency greater than~3$\fc$ focused on investigation the dependence of severity of voltage fluctuation on their shape in the case of ``clipped cosine" supply voltage distortion (``clipped cosine" distortion is a good approximation of states occurring in LV networks). On the other hand, voltage fluctuations in  power grids are modelled as amplitude modulation~(AM) without a suppressed carrier wave~\cite{b25,b26,b27}, which is a good approximation of voltage fluctuations in a stiff power grid (in the case of island operation, the phase/frequency variation should also be considered, i.e., voltage fluctuations should be modelled as amplitude and frequency/phase modulation~(AM-FM/PM)~\cite{b28,b29}). For the adopted model of voltage fluctuations, ``clipped cosine" supply voltage distortion is adopted as the carrier signal, and arbitrary signals associated with the influence of potential voltage fluctuations sources are adopted as the modulating signals. For the adopted model, numerical simulation studies and experimental laboratory tests were carried out, on the basis of which the dependence of severity of voltage fluctuations on their shape are determined.
	
\section{Adopted model of power quality disturbances}

As the model of power quality disturbances used in laboratory tests and in numerical simulation studies is adopted the signal (AM modulation without suppressed carrier wave) described by the relationship: 
\begin{equation}\label{eq1}
	u_{IN}\left( t\right) = \left( 1 + \left( \frac{\Delta U}{U} \right) \frac{1}{2 \cdot 100} u_{\text{mod}}\left( t\right)  \right) \cdot u_c\left( t\right), 
\end{equation}
where: $\umodt$ is the normalized modulating signal associated with the influence of voltage fluctuations sources, $\dUU$ is the modulation depth related to the nature of power grids and disturbing loads expressed in \%, $\uct$ is the supply voltage in the power grid prior to the appearance voltage fluctuations sources (carrier signal).

The research consider a ``clipped cosine" signal with three values of $\mc$ as the carrier signal, where $\mc$ is the clipping level described by the relationship:
\begin{equation}\label{eq2}
	\mc = \frac{M_c}{M},
\end{equation}
where: $M_c$ and $M$ are the signal amplitude after and before clipping, respectively. The selected $\mc$ values are:
\begin{itemize}
	\item $\mc = 1$ -- rated purely sinusoidal supply voltage;
	\item $\mc = 0.8$ -- typical distorted voltage in the power grid caused by electronic devices (by input stages of switching power supplies) for which total harmonic distortion (THD) ratio is approx.~8\% (the limit value for LV networks~\cite{b30}) -- it is worth noting that usually in real LV networks the value~of~$\mc \in [0.8; 1.0]$;
	\item $\mc = 0.1$ -- highly distorted voltage in the power grid, typical for example for island operation supplied with some converters~\cite{b31,b32,b33}, or for UPS~\cite{b33}, or for resonance interactions in the power grid causing an increase in the contribution of higher harmonics; for the assumed $\mc$~value, the shape of waveform is similar to a square wave and THD ratio is approx.~43\%.   
\end{itemize}

\begin{figure}[H]\centering
	\includegraphics[width=\columnwidth]{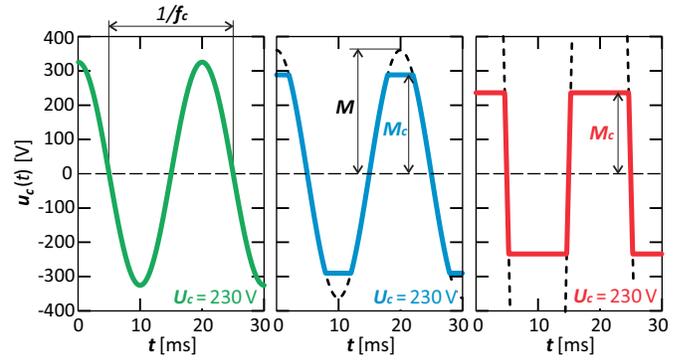}
	\caption{Assumed carrier signals with~$\mc$ equal to 1.0~(left), 0.8~(centre), 0.1~(right), respectively.}
	\label{fig1}
\end{figure}

\noindent
In the research, for every assumed carrier signals, the fundamental frequency~$\fc$ was 50\unit{Hz} (rated power frequency~\cite{b30}) and the rms value~$U_c$ was 230\unit{V} (rated rms value in LV networks~\cite{b30}). Assumed carrier signals (supply voltage before voltage fluctuations) are shown in Fig.~\ref{fig1}.

Four types of signals are considered as the modulating signal:
\begin{itemize}
	\item sinusoidal signal;
	\item triangular signal;
	\item trapezoidal signal (duration of the rising edge and falling edge equal~to~1/4$T_m$, where $T_m$ is the fundamental period of modulating signal) -- this shape of voltage fluctuations can be caused by loads equipped with a soft starter system~\cite{b35,b36};
	\item rectangular signal -- associated with the influence of loads causing step voltage changes, which are probably the most common in the power grid nowadays~\cite{b36,b37,b38}.
\end{itemize}

\begin{figure}[H]\centering
	\includegraphics[width=.95\columnwidth]{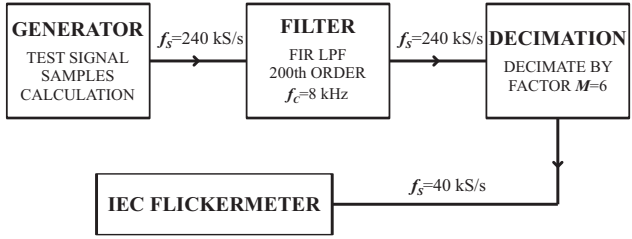}
	\caption{The block diagram of signal chain in MATLAB, where LPF is a low-pass filter.}
	\label{fig1a}
\end{figure}

\noindent
It is worth noting that the real (resultant) shapes of voltage fluctuations probably have more complex waveforms, although assumed arbitrary test signals make it possible to observe some tendencies~\cite{b7} and are a good starting point for further research. The studies were carried out in two stages. In both stages, for the generated test signal described by~(\ref{eq1}), the short-term flicker indicator~$\Pst$~\cite{b39} was determined and recorded, and which was adopted in the research as the reference value determining voltage fluctuation severity. It is worth noting that despite limitations of this measure of voltage fluctuation severity, it is currently used in many countries and allows a one-parameter and numerical assessment of one potential effects of voltage fluctuations. In numerical simulation studies, test signal generation, bandwidth limitation (using a low-pass FIR filter of order 200 and a cut-off frequency of~8\unit{kHz}), downsampling and recording $\Pst$~values (using the IEC flickermeter implementation~\cite{b39a}) for the assumed modulated signal with a limited bandwidth was performed in MATLAB (see Fig.~\ref{fig1a}). 

\begin{figure}[htbp]\centering
	\includegraphics[width=.7\columnwidth]{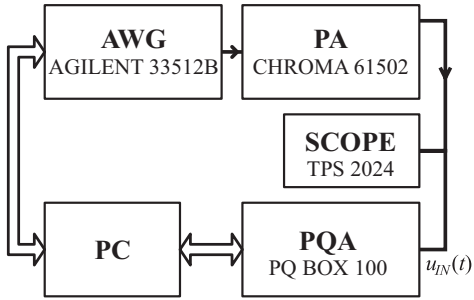}
	\caption{The laboratory setup, where: AWG is a arbitrary waveform generator~\cite{b41}, PA is power amplifier~\cite{b42}, PQA is class A power quality analyser~\cite{b43} with built-in the IEC flickermeter~\cite{b44,b45}.}
	\label{fig2}
\end{figure}

\noindent
In experimental laboratory tests, test signal generation and recording $\Pst$~values for it were performed in the laboratory setup shown in Fig.~\ref{fig2}. It is worth noting that $\Pst$~values for individual test signals were recorded for 20 minutes (two 10-minute intervals) and the first 10-minute interval related to transient states~\cite{b40} was rejected. Hence, considering the time-consuming of experimental studies, they were carried out only for stage I of the research for $\dUU = 5\%$ for all assumed carrier and modulating signals. It is worth noting that the results of simulation and experimental tests are consistent, so without losing generality, it can be assumed that the other simulations (for which experimental studies are not carried out) are correct. The inconsistency occurs only for $\Pst<0.1$, which results from the imperfection of the signal chain in the real flickermeter related to, e.g., the resolution of instrument, implementation of particular filters.

In the first stage of the research, individual carrier signals were modulated with selected modulating signals with a modulation depth~$\dUU$ equal to 1\%, 5\% and 10\%, respectively, and with a modulating frequency~$\fm \in$~[0.01;~1050]\unit{Hz}. In the second stage of the research, carrier signals were modulated with selected modulating signals with modulating frequency~$\fm$ equal to 208.8\unit{Hz} and 1008.8z\unit{Hz} and modulation depth~$\dUU \in$[0.1;~20]\%. It is worth noting that for the modulating frequency~$\fm$ at the level of\unit{kHz}, the modulation depth~$\dUU$ at the level of 10\% causes variations of the rms values determined for the fundamental period every half of the fundamental period (the fundamental period determined with the use of zero cross detection) at the level of 0.1\%, so in fact for large $\fm$~values, the modulation depth~$\dUU$ in the voltage fluctuations model can be much greater than 10\%.

\section{Research results and discussion}

\begin{figure}[htbp]\centering
	\includegraphics[width=.95\columnwidth]{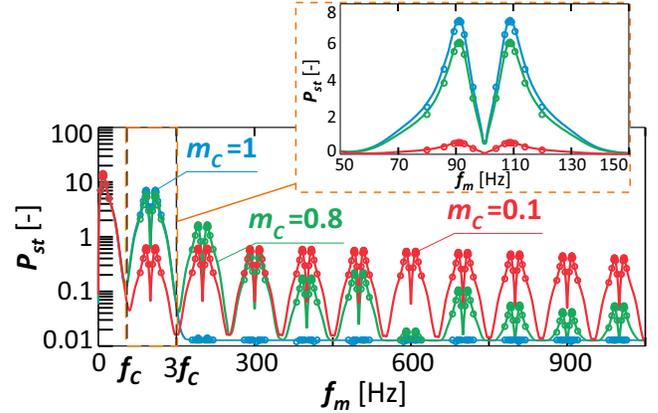}
	\caption{The $\Pst = \textbf{f} \left( \fm, \dUU = 5\% = \const \right)$ characteristics for the sinusoidal modulating signal, where $\mc$ is 1.0 (blue), 0.8 (green), 0.1 (red), respectively (line -- simulation results, markers -- laboratory test results).}
	\label{fig3}
\end{figure}

As a reminder, Fig.~\ref{fig3} shows a comparison of characteristics $\Pst=\textbf{f} \left( \fm, \dUU = 5\% = \const \right) $ for the sinusoidal modulating signal and for every assumed carrier signals. 

\begin{figure}[H]\centering
	\includegraphics[width=.95\columnwidth]{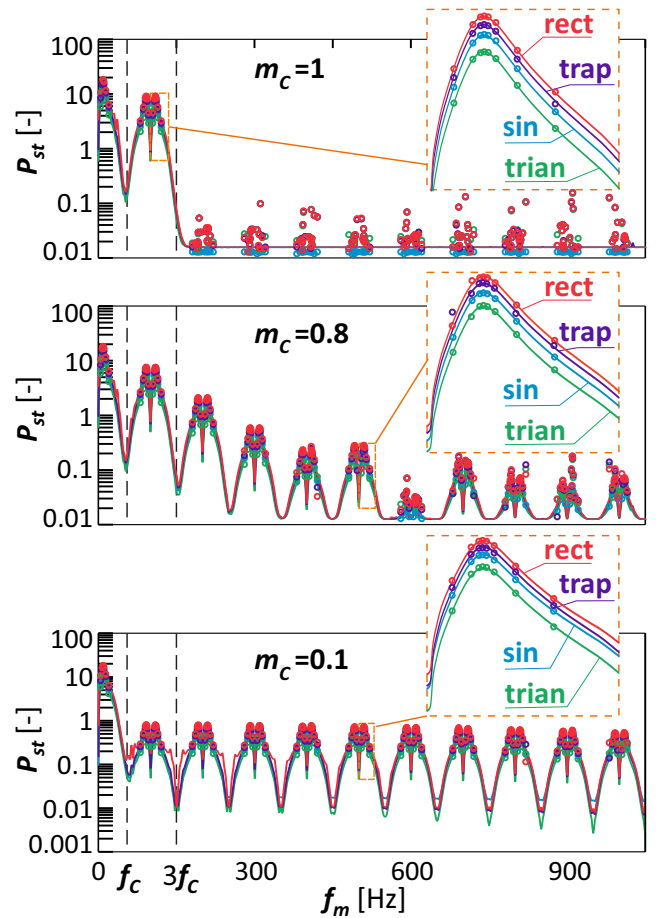}
	\caption{The $\Pst = \textbf{f} \left( \fm, \dUU = 5\% = \const \right)$ characteristics for sinusoidal (blue), triangular (green), trapezoidal (purple), rectangular (red) modulating signal, where $\mc$ is 1.0 (top), 0.8 (middle), 0.1 (bottom), respectively (line -- simulation results, markers -- laboratory test results).}
	\label{fig4}
\end{figure}

\noindent
In Fig.~\ref{fig3} it can be seen that for the sinusoidal carrier signal (supply voltage), values of~$\Pst > 0$ occur only for $\fm < 3 \fc$. On the other hand, when the carrier signal is distorted, values of~$\Pst > 0$ can occur for $\fm > 3 \fc$. It is worth noting that, depending on the distortion level, values of~$\Pst > 0$ occur for different intervals of~$\fm$. When $\mc \to 0$ (possible situation for island operation or selected emergency states of power grids), then values of~$\Pst > 0$ can occur for ever greater $\fm$~values. The dependence of values of~$\Pst > 0$ occurrence for individual intervals of~$\fm$ is non-linear, non-monotonic and difficult to precisely define, as presented in~\cite{b5}. It is worth noting that the voltage distortion in power grids (carrier signal) does not affect the~$\Pst$~measurement result for~$\fm$ in the range [0;~$\fc$]\unit{Hz}. For~$\fm$ in the range ($\fc$;~3$\fc$]\unit{Hz}, the influence of voltage distortion on the~$\Pst$~measurement result is noticeable, and in the range for~$\fm > 3 \fc$, the voltage distortion can cause possibility occurrence values of~$\Pst > 0$ (experimentally verified visible obnoxious flicker -- luminance variation).
\newline

\begin{figure}[H]\centering
	\includegraphics[width=.95\columnwidth]{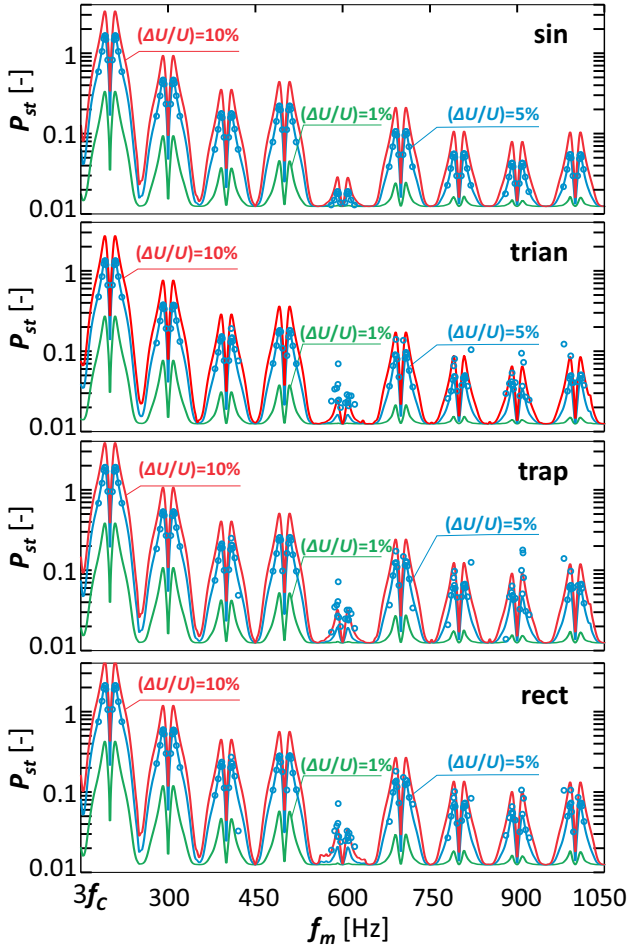}
	\caption{The $\Pst = \textbf{f} \left( \fm, \dUU \in \left\lbrace 1, 5, 10 \right\rbrace \% = \const \right)$ characteristics for sinusoidal (first from top), triangular (second from top), trapezoidal (third from top), rectangular (fourth from top) modulating signal, where $\mc$ for carrier signal is 0.8 (line -- simulation results, markers -- laboratory test results).}
	\label{fig5}
\end{figure}

Fig.~\ref{fig4} shows a comparison of characteristics $\Pst = \textbf{f} \left( \fm, \dUU = 5\% = \const \right) $ for every assumed modulating signals and carrier signals. On the basis of characteristics, it can be seen that a some tendency for the dependence of severity of voltage fluctuations on their shape for $\fm > 3 \fc$ is maintained as for the range $\fm < 3 \fc$, i.e., with the same modulation depth, the most obnoxious are rectangular, trapezoidal, sinusoidal and triangular voltage fluctuations, respectively. Assuming that for any $\fm$~values there is a tendency to increase severity of voltage fluctuations depending on their shape, it is possible to generalized properties for any~$\fm$, based on the results for $\fm$~values in the range [0;~3$\fc$]\unit{Hz}. However, this is a hypothesis that requires analytical proof. In the next part, the paper focuses mainly on the voltage fluctuation with a frequency $\fm > 3 \fc$ ($\fc$ = 50\unit{Hz}), i.e., considered characteristics are presented in the range $\fm \in $[150;~1050]\unit{Hz}. In this range, the effect of changing the modulation depth~$\dUU$ and the modulating frequency~$\fm$ on $\Pst$~measurement results is determined for assumed shapes of modulating signals.
\newline

\begin{figure}[H]\centering
	\includegraphics[width=.95\columnwidth]{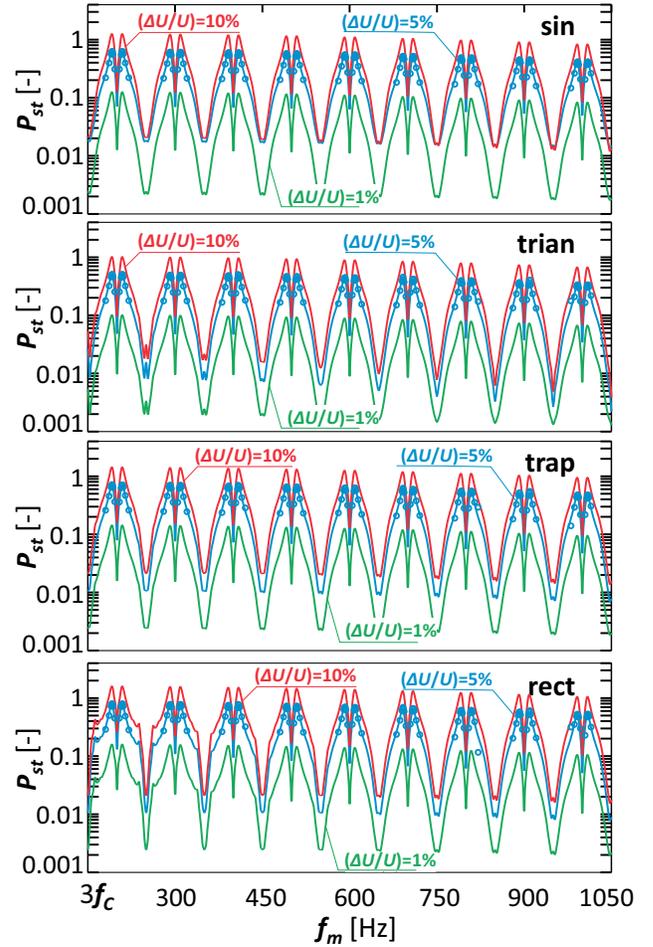}
	\caption{The $\Pst = \textbf{f} \left( \fm, \dUU \in \left\lbrace 1, 5, 10 \right\rbrace \% = \const \right)$ characteristics for sinusoidal (first from top), triangular (second from top), trapezoidal (third from top), rectangular (fourth from top) modulating signal, where $\mc$ for carrier signal is 0.1 (line -- simulation results, markers -- laboratory test results).}
	\label{fig6}
\end{figure}

Fig.~\ref{fig5} and Fig.~\ref{fig6} show a comparison of characteristics $\Pst = \textbf{f} \left( \fm, \dUU \in \left\lbrace 1, 5, 10 \right\rbrace  \% = \const \right) $ for every assumed modulating signals and for the assumed carrier signals with $\mc$~equal to 0.8 (Fig.~\ref{fig5}) and 0.1 (Fig.~\ref{fig6}) respectively. Based on the characteristics, it can be seen that for individual values of~$\fm$, regardless of the shape of modulating signal, the same tendency to increase $\Pst$~value with increasing modulation depth~$\dUU$ occurs.

\begin{figure}[htbp]\centering
	\includegraphics[width=.95\columnwidth]{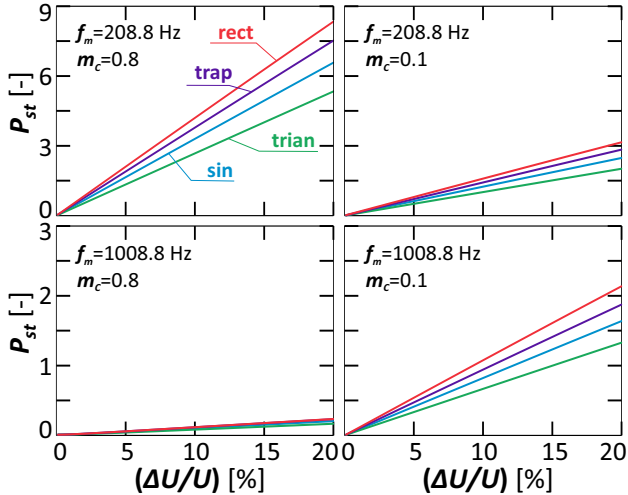}
	\caption{The $ \Pst = \textbf{f} \left( \fm \in \left\lbrace 208.8, 1008.8\right\rbrace \unit{Hz} = \const, \dUU \right)  $ characteristics for sinusoidal (blue), triangular (green), trapezoidal (purple), rectangular (red) modulating signal, where $\mc$ for the carrier signal is 0.8 (left) and 0.1(right), respectively (line -- simulation results).}
	\label{fig7}
\end{figure}

Fig.~\ref{fig7} shows a comparison of characteristics $\Pst = \textbf{f} \left( \fm \in \left\lbrace 208.8, 1008.8\right\rbrace \unit{Hz} = \const, \dUU \right) $ for every assumed modulating signals and for the assumed carrier signals with $\mc$~equal to 0.8 and 0.1, respectively. On the basis of the characteristics, it can be seen that for the assumed $\fm$~values, the dependence of $\Pst$~value on modulation depth~$\dUU$, regardless of the shape of modulating signal, is approximately linear. It is worth noting that the slope of characteristic depends on the modulating signal shape and the carrier signal shape. Considering the non-linearity and non-monotonicity of the dependence of $\Pst$~value on $\fm$~value for different $\mc$~value~\cite{b5} (see Fig.~\ref{fig7} -- for $\mc = 0.8$ greater $\Pst$~value is achieved for $\fm = 208.8\unit{Hz} $ at the same modulation depths~$\dUU$ than for $\mc = 0.1$, and for $\fm = 1008.8\unit{Hz} $ the relationship is inverse), it is not possible to explicitly define the dependence of slope of characteristic $\Pst = \textbf{f} \left( \fm = \const, \dUU \right) $ on shape of carrier signal (the supply voltage prior to the occurrence of voltage fluctuations sources) and on shape of modulating signal (specific for individual voltage fluctuations sources).

\section{Conclusion}

The paper presents the IEC flickermeter measurement results for assumed distorted modulating signals with simultaneous ``clipped cosine" distortion of the carrier signal for selected clipping levels~$\mc$. On the basis of results obtained for individual test signals, it can be seen that if the supply voltage (carrier signal) is distorted, then for $\fm > 3\fc$, obnoxious voltage fluctuations ($\Pst >> 0$) can occur. Moreover, along with the increase in distortion of supply voltage (carrier signal), the voltage fluctuations severity can increase for ever greater $\fm$~values. Also for $\fm > 3\fc$, analogous tendencies in the dependence of voltage fluctuation shape are noticeable as for $\fm$~values in the range [0;~3$\fc$]\unit{Hz} (also in the conventional situation, where a sinusoidal carrier signal is assumed). It is worth noting that the presented preliminary research results are a good start for further research aimed at determining the potential voltage fluctuation severity caused by specific power electronic devices. In order to precisely determine the voltage fluctuation severity from the level of distortion of modulating signal and the level of distortion of carrier signal, it is necessary to prepare an analytical IEC flickermeter model, and then carry out experimental verification, considering modulating frequency at the level of\unit{kHz} associated with commonly used power electronics devices (e.g., inverters), which is a further research target.

\vspace{12pt}

\end{document}